\documentclass[aps,prl,twocolumn,showpacs,groupedaddress,floatfix]{revtex4}  
\usepackage{graphicx}  
\usepackage{dcolumn}   
\usepackage{bm}        
\usepackage{amssymb}   

\newcommand{\scriptR}{$\xi$} 
\newcommand{\MET}{\mbox{$\protect \raisebox{.3ex}{$\not$}E_T$}}

\begin{document}


\hspace{5.2in} \mbox{Fermilab-Pub-06/167-E}

\title{Search for a heavy resonance decaying into a $Z+$jet final state \\
        in $p\bar p$ collisions at $\sqrt{s}= 1.96$ TeV using the D0 detector}
%
\author{                                                                      
V.M.~Abazov,$^{36}$                                                           
B.~Abbott,$^{76}$                                                             
M.~Abolins,$^{66}$                                                            
B.S.~Acharya,$^{29}$                                                          
M.~Adams,$^{52}$                                                              
T.~Adams,$^{50}$                                                              
M.~Agelou,$^{18}$                                                             
J.-L.~Agram,$^{19}$                                                           
S.H.~Ahn,$^{31}$                                                              
M.~Ahsan,$^{60}$
% ...
\\                                                                            
\vskip 0.30cm                                                                 
\centerline{(D\O\ Collaboration)}                                             
\vskip 0.30cm                                                                 
}                                                                             
\affiliation{                                                                 
\centerline{$^{1}$Universidad de Buenos Aires, Buenos Aires, Argentina}       
\centerline{$^{2}$LAFEX, Centro Brasileiro de Pesquisas F{\'\i}sicas,         
                  Rio de Janeiro, Brazil}                                     
\centerline{$^{3}$Universidade do Estado do Rio de Janeiro,                   
                  Rio de Janeiro, Brazil}                                     
\centerline{$^{4}$Instituto de F\'{\i}sica Te\'orica, Universidade            
                  Estadual Paulista, S\~ao Paulo, Brazil}                     
\centerline{$^{5}$University of Alberta, Edmonton, Alberta, Canada,           
                  Simon Fraser University, Burnaby, British Columbia, Canada,}
\centerline{York University, Toronto, Ontario, Canada, and                    
                  McGill University, Montreal, Quebec, Canada}                
\centerline{$^{6}$Institute of High Energy Physics, Beijing,                  
                  People's Republic of China}                                 
\centerline{$^{7}$University of Science and Technology of China, Hefei,       
                  People's Republic of China}                                 
\centerline{$^{8}$Universidad de los Andes, Bogot\'{a}, Colombia}             
\centerline{$^{9}$Center for Particle Physics, Charles University,            
                  Prague, Czech Republic}                                     
\centerline{$^{10}$Czech Technical University, Prague, Czech Republic}        
\centerline{$^{11}$Center for Particle Physics, Institute of Physics,         
                   Academy of Sciences of the Czech Republic,                 
                   Prague, Czech Republic}                                    
\centerline{$^{12}$Universidad San Francisco de Quito, Quito, Ecuador}        
\centerline{$^{13}$Laboratoire de Physique Corpusculaire, IN2P3-CNRS,         
                   Universit\'e Blaise Pascal, Clermont-Ferrand, France}      
\centerline{$^{14}$Laboratoire de Physique Subatomique et de Cosmologie,      
                   IN2P3-CNRS, Universite de Grenoble 1, Grenoble, France}    
\centerline{$^{15}$CPPM, IN2P3-CNRS, Universit\'e de la M\'editerran\'ee,     
                   Marseille, France}                                         
\centerline{$^{16}$IN2P3-CNRS, Laboratoire de l'Acc\'el\'erateur              
                   Lin\'eaire, Orsay, France}                                 
\centerline{$^{17}$LPNHE, IN2P3-CNRS, Universit\'es Paris VI and VII,         
                   Paris, France}                                             
\centerline{$^{18}$DAPNIA/Service de Physique des Particules, CEA, Saclay,    
                   France}                                                    
\centerline{$^{19}$IReS, IN2P3-CNRS, Universit\'e Louis Pasteur, Strasbourg,  
                    France, and Universit\'e de Haute Alsace,                 
                    Mulhouse, France}                                         
\centerline{$^{20}$Institut de Physique Nucl\'eaire de Lyon, IN2P3-CNRS,      
                   Universit\'e Claude Bernard, Villeurbanne, France}         
\centerline{$^{21}$III. Physikalisches Institut A, RWTH Aachen,               
                   Aachen, Germany}                                           
\centerline{$^{22}$Physikalisches Institut, Universit{\"a}t Bonn,             
                   Bonn, Germany}                                             
\centerline{$^{23}$Physikalisches Institut, Universit{\"a}t Freiburg,         
                   Freiburg, Germany}                                         
\centerline{$^{24}$Institut f{\"u}r Physik, Universit{\"a}t Mainz,            
                   Mainz, Germany}                                            
\centerline{$^{25}$Ludwig-Maximilians-Universit{\"a}t M{\"u}nchen,            
                   M{\"u}nchen, Germany}                                      
\centerline{$^{26}$Fachbereich Physik, University of Wuppertal,               
                   Wuppertal, Germany}                                        
\centerline{$^{27}$Panjab University, Chandigarh, India}                      
\centerline{$^{28}$Delhi University, Delhi, India}                            
\centerline{$^{29}$Tata Institute of Fundamental Research, Mumbai, India}     
\centerline{$^{30}$University College Dublin, Dublin, Ireland}                
\centerline{$^{31}$Korea Detector Laboratory, Korea University,               
                   Seoul, Korea}                                              
\centerline{$^{32}$SungKyunKwan University, Suwon, Korea}                     
\centerline{$^{33}$CINVESTAV, Mexico City, Mexico}                            
\centerline{$^{34}$FOM-Institute NIKHEF and University of                     
                   Amsterdam/NIKHEF, Amsterdam, The Netherlands}              
\centerline{$^{35}$Radboud University Nijmegen/NIKHEF, Nijmegen, The          
                  Netherlands}                                                
\centerline{$^{36}$Joint Institute for Nuclear Research, Dubna, Russia}       
\centerline{$^{37}$Institute for Theoretical and Experimental Physics,        
                   Moscow, Russia}                                            
\centerline{$^{38}$Moscow State University, Moscow, Russia}                   
\centerline{$^{39}$Institute for High Energy Physics, Protvino, Russia}       
\centerline{$^{40}$Petersburg Nuclear Physics Institute,                      
                   St. Petersburg, Russia}                                    
\centerline{$^{41}$Lund University, Lund, Sweden, Royal Institute of          
                   Technology and Stockholm University, Stockholm,            
                   Sweden, and}                                               
\centerline{Uppsala University, Uppsala, Sweden}                              
\centerline{$^{42}$Physik Institut der Universit{\"a}t Z{\"u}rich,            
                   Z{\"u}rich, Switzerland}                                   
\centerline{$^{43}$Lancaster University, Lancaster, United Kingdom}           
\centerline{$^{44}$Imperial College, London, United Kingdom}                  
\centerline{$^{45}$University of Manchester, Manchester, United Kingdom}      
\centerline{$^{46}$University of Arizona, Tucson, Arizona 85721, USA}         
\centerline{$^{47}$Lawrence Berkeley National Laboratory and University of    
                   California, Berkeley, California 94720, USA}               
\centerline{$^{48}$California State University, Fresno, California 93740, USA}
\centerline{$^{49}$University of California, Riverside, California 92521, USA}
\centerline{$^{50}$Florida State University, Tallahassee, Florida 32306, USA} 
\centerline{$^{51}$Fermi National Accelerator Laboratory,                     
            Batavia, Illinois 60510, USA}                                     
\centerline{$^{52}$University of Illinois at Chicago,                         
            Chicago, Illinois 60607, USA}                                     
\centerline{$^{53}$Northern Illinois University, DeKalb, Illinois 60115, USA} 
\centerline{$^{54}$Northwestern University, Evanston, Illinois 60208, USA}    
\centerline{$^{55}$Indiana University, Bloomington, Indiana 47405, USA}       
\centerline{$^{56}$University of Notre Dame, Notre Dame, Indiana 46556, USA}  
\centerline{$^{57}$Purdue University Calumet, Hammond, Indiana 46323, USA}    
\centerline{$^{58}$Iowa State University, Ames, Iowa 50011, USA}              
\centerline{$^{59}$University of Kansas, Lawrence, Kansas 66045, USA}         
\centerline{$^{60}$Kansas State University, Manhattan, Kansas 66506, USA}     
\centerline{$^{61}$Louisiana Tech University, Ruston, Louisiana 71272, USA}   
\centerline{$^{62}$University of Maryland, College Park, Maryland 20742, USA} 
\centerline{$^{63}$Boston University, Boston, Massachusetts 02215, USA}       
\centerline{$^{64}$Northeastern University, Boston, Massachusetts 02115, USA} 
\centerline{$^{65}$University of Michigan, Ann Arbor, Michigan 48109, USA}    
\centerline{$^{66}$Michigan State University,                                 
            East Lansing, Michigan 48824, USA}                                
\centerline{$^{67}$University of Mississippi,                                 
            University, Mississippi 38677, USA}                               
\centerline{$^{68}$University of Nebraska, Lincoln, Nebraska 68588, USA}      
\centerline{$^{69}$Princeton University, Princeton, New Jersey 08544, USA}    
\centerline{$^{70}$State University of New York, Buffalo, New York 14260, USA}
\centerline{$^{71}$Columbia University, New York, New York 10027, USA}        
\centerline{$^{72}$University of Rochester, Rochester, New York 14627, USA}   
\centerline{$^{73}$State University of New York,                              
            Stony Brook, New York 11794, USA}                                 
\centerline{$^{74}$Brookhaven National Laboratory, Upton, New York 11973, USA}
\centerline{$^{75}$Langston University, Langston, Oklahoma 73050, USA}        
\centerline{$^{76}$University of Oklahoma, Norman, Oklahoma 73019, USA}       
\centerline{$^{77}$Oklahoma State University, Stillwater, Oklahoma 74078, USA}
\centerline{$^{78}$Brown University, Providence, Rhode Island 02912, USA}     
\centerline{$^{79}$University of Texas, Arlington, Texas 76019, USA}          
\centerline{$^{80}$Southern Methodist University, Dallas, Texas 75275, USA}   
\centerline{$^{81}$Rice University, Houston, Texas 77005, USA}                
\centerline{$^{82}$University of Virginia, Charlottesville,                   
            Virginia 22901, USA}                                              
\centerline{$^{83}$University of Washington, Seattle, Washington 98195, USA}  
}                                                                             
\date{June 8, 2006}

\begin{abstract}
We have searched for a heavy resonance decaying
into a $Z+$jet final state in $p\bar p$ collisions at a
center of mass energy of $1.96$ TeV at the Fermilab Tevatron collider
using the D0 detector. No indication for 
such a resonance was found in a data sample corresponding to 
an integrated luminosity of 370 pb$^{-1}$. 
We set 
upper limits on the cross section times branching fraction 
for heavy resonance production
at the 95\% C.L.
as a function of the resonance mass and width.
The limits are interpreted within  the framework of a
specific model of excited quark production.
\end{abstract}

\maketitle 


Heavy resonances decaying into a quark and a gauge boson may signal
the existence of excited quarks and thereby indicate 
quark substructure \cite{Baur90}. Searches for excited quarks
have been carried out in the past using 
dijet \cite{UA2dij, CDFdij, D0dij},
photon$+$jet, and $W+$jet \cite{CDFWgamj}  final states. In the 
analysis described here, we searched for resonances in the $Z+$jet
channel, where the $Z$ boson is detected via its $Z\to e^+e^-$
decay mode. This signature is practically free of instrumental background.
However, it suffers from the low branching fraction
(3.36\%)
of the $Z\to e^+e^-$ decay channel. The high
luminosity delivered by the Fermilab Tevatron collider in Run II makes it possible
to present results on 
this final state for the first time.

For the production and decay of a resonance, we considered the
model \cite{Baur90}
implemented in {\sc pythia 6.202} \cite{pythia}. 
Here, a quark (antiquark) and a gluon from the colliding proton
and antiproton form a resonance, $q^*$, which subsequently decays
into a $Z$ boson and a quark: $q^*\to q+Z$.
The model  
has two free parameters,
$M_{q^*}$, the mass of the resonance, and $\Lambda$, the
compositeness scale.  They determine
the production cross section and the natural width of the resonance.
The latter scales as $1/$\scriptR$^2$, where \scriptR\ $ = \Lambda/M_{q^*}$.

The Run II D0 detector \cite{d0det} consists of several layered subdetectors. 
For the present analysis, the most relevant parts are 
the liquid-argon/uranium calorimeter \cite{d0calo} and the central tracking system. 
The calorimeter, 
divided into electromagnetic and hadronic sections, 
has 
a 
granularity of $\Delta\eta\times\Delta\phi = 0.1 \times 0.1$,
where 
$\eta$ is the pseudorapidity ($\eta = -\ln\left[\tan (\theta/2)\right]$ with
$\theta$ being the polar angle  
measured from the geometrical center of the detector with
respect to the proton beam direction)
and $\phi$ is the azimuthal angle. 
The third innermost layer, in which  the largest electromagnetic energy deposition
is expected, has a finer granularity of
$\Delta\eta\times\Delta\phi = 0.05 \times 0.05$.
The central calorimeter 
covers $|\eta| \le 1.1$, and the two end calorimeters 
extend coverage 
to $|\eta|\approx 4.5$. 
The tracking system consists of a 
silicon microstrip tracker  and a central fiber tracker, 
both located within a 2~T superconducting solenoidal 
magnet, with designs optimized for tracking and 
vertexing at pseudorapidities $|\eta|<3$ and $|\eta|<2$, respectively.

The data used in this analysis were collected between April 2002 and
August 2004, with an integrated luminosity of 
$370$ pb$^{-1}$.
The selected events were required to pass at least one
of several single- or di-electron triggers.
The efficiency of the trigger
was measured with data
and found to reach a plateau of $\varepsilon_{\text{trig}}= 0.982\pm 0.011$ 
for events satisfying the final event selection criteria.

Offline event selection was based on run quality, event properties, 
and electron and jet identification criteria.
Events were required to have a reconstructed vertex with a longitudinal
position within 60 cm of the detector center.
Electrons were reconstructed from electromagnetic (EM) clusters in the
calorimeter using a 
cone algorithm. 
The reconstructed electron candidates were required to 
satisfy either  
$|\eta | \le 1.1$ or $ 1.5 < |\eta | \le 2.5 $.
Electron pairs  with $p_T^{e1}\ge 30$ GeV and $p_T^{e2}\ge 25$ GeV 
in the event were used to reconstruct
the $Z$ boson candidate. The electron pair was required to have an
invariant mass $M_{ee}$ 
near the $Z$ boson mass,
$80<M_{ee}<102$ GeV.

\begin{figure}
\includegraphics[scale=0.4]{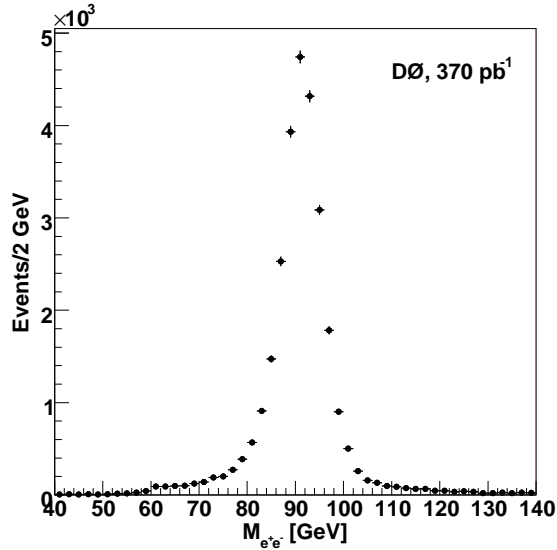}
\caption{ \label{fig:ElecSel} The invariant mass of the
two selected electrons in the data events. 
}
\end{figure}

To reduce background contamination, mainly from jets 
misidentified as
electrons,
the EM clusters were required to pass three quality criteria based on shower 
profile: 
$(i)$ the ratio of the 
energy deposited in the electromagnetic part of the calorimeter 
to the total shower energy 
had to exceed 0.9; $(ii)$ the lateral and longitudinal shapes
of the energy cluster had to be consistent with those of an electron;
and $(iii)$ the electron had to be isolated from other energy deposits 
in the calorimeter with isolation fraction $f_{iso}<0.15$. 
The isolation fraction is defined as 
$f_{iso}=\left[ E(0.4)-E_{EM}(0.2)\right]/E_{EM}(0.2)$, where 
$E(R_{\text {cone}})$ and $E_{EM}(R_{\text {cone}})$ are the total and the EM
energy, respectively, deposited within a cone of radius 
$R_{\text {cone}} = \sqrt{(\Delta\eta)^2+(\Delta\phi)^2}$
centered around the electron. 
Additionally, at least one
of the electrons was required 
to have a spatially close track with a
momentum consistent with the EM shower energy.
A total of  24,734 events passed these criteria.
In Fig.~\ref{fig:ElecSel},  the distribution of the invariant mass,
$M_{ee}$,  of the two selected electrons is shown.
A very clean, almost background-free $Z$ boson signal is evident.

Jets were reconstructed using the ``Run II cone algorithm''~\cite{RunIIcone}
which combines 
cell energies
within a cone of radius $R_{\text {cone}} =  0.5$.
Spurious jets from isolated noisy calorimeter cells were supressed
by cuts on the jet shape 
and by requiring that the
charged tracks associated with the jet
had to carry  a minimum fraction
of the jet transverse energy.
The transverse momentum of each jet
was corrected  for offsets due to the underlying event,
multiple $p\bar p$ interactions and noise, out-of-cone showering,
and the detector energy response as determined from the 
transverse energy balance of photon$+$jet events.
Jets were required to have  $p_T>$ 20 GeV and $|\eta |<$ 2.5
and to not 
overlap
with any of the
reconstructed EM objects within a distance of
0.4 in $(\eta,\phi)$ space.
Requiring one or more jets with these selection criteria, 
2,417 data events remain.

We have considered two kinds of instrumental backgrounds where hadronic
jets are misreconstructed as EM clusters and mimic $Z$ boson events.
A background from genuine QCD multi-jet production arises when both of the EM objects are hadronic
jets that fluctuate to electromagnetic final states.  This background has been estimated to be
$(0.56 \pm 0.02)$\% of the signal
in the mass region of  $80<M_{ee}<102$ GeV
as calculated by comparing the $M_{ee}$ mass
distribution of the selected events with a distribution that required inverted shower shape
criteria and the absence of matching tracks.
The other source of background is $W\to e\nu+$jets
events where a hadronic jet 
is misidentified as an electron. 
These events are characterized by significant missing transverse
energy (\MET), and should also appear in the data sample where only one of the EM
objects has a matched track. From comparison of the 
\MET\  distribution
of these events with that where both electrons do have matched tracks,
we estimate that this  background is an order of magnitude
less than the QCD background. 

The main standard model (SM) background to the excited quark signal is
inclusive $Z/\gamma^*\to e^+e^-$ pair production which has been simulated with {\sc pythia} 
using the CTEQ5L \cite{PDFqstar} parton distribution functions (PDFs).
In order to enhance
the statistics for events 
where the invariant mass of the $Z$ boson and the leading jet, 
$M_{Zj1}$, 
is high,
in addition to the so-called $2\to 1$ process, 
we have also generated 
events including 
matrix elements of first order in 
$\alpha_s$
($2\to 2$ process) with different thresholds
of $M_{Zq}$, the invariant mass of the $Z$ boson and the 
accompanying parton
in the final state. A minimum value of 30 GeV for
$p_{Tp}$, the transverse momentum of the parton in the
$2\to 2$ collision, has been set in order
to avoid collinear divergences. 
The leading jet $p_T$ distribution 
has a mean and an RMS value of 106 GeV and 27 GeV, respectively, 
at the lowest resonance mass investigated and 
after the final selection, therefore
the  $p_{Tp}$ cut does not affect the analysis.
The shape of the $M_{Zj1}$ distribution has been compared with that obtained with  
the {\sc alpgen} program \cite{Alpgen}
and there is reasonable agreement between them.
Any differences in the background level 
have been taken into account as a systematic uncertainty.

Signal events were generated with {\sc pythia} 
using the CTEQ5L PDFs 
for the following resonance mass values:
$M_{q^*} =$ 300, 400, 500, 600 and 700 GeV with \scriptR\  $ = \Lambda/M_{q^*} = 1$.
For each mass, except for the lowest one, we also generated events
with \scriptR\  $=  0.3$, 0.5, 0.7, in order to vary the natural width
of the resonance, $\Gamma_{q^*}$. 
The form factors associated with the interaction of the quarks
with the SM gauge bosons were set to unity.
The MC events were passed through
the same reconstruction software
and selection criteria as the data.
The events have been used to estimate the geometrical acceptance and
jet and electron identification
efficiencies. 
The combined acceptance times efficiencies are listed in Table~\ref{tab:results}.
The resolution of 
$M_{Zj1}$ has been found to be $\approx$ 9\%. 

\begin{figure}
\includegraphics[scale=0.4]{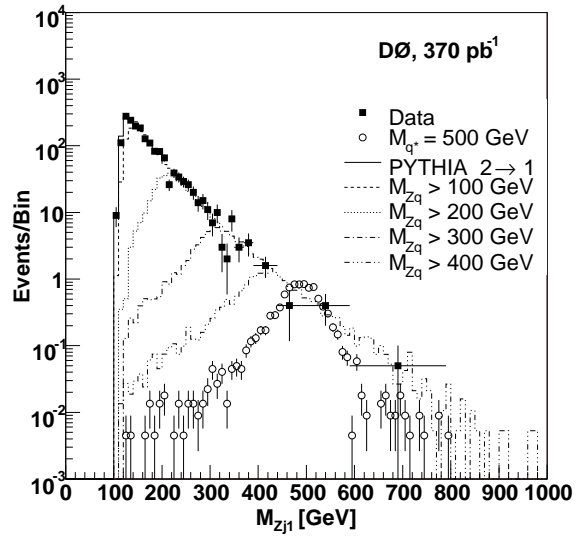}
\caption{ \label{fig:MZj1DaMCcompnocut} Invariant mass
distribution of the $Z$ boson and the leading jet, $M_{Zj1}$.
The data are shown by the full squares with error bars. 
The actual number of events in a bin is the product of the
plotted value and the bin width measured in 10 GeV units.
The SM backgrounds 
generated with {\sc pythia} are shown in the
histograms: $2\to 1$ without threshold (solid line), 
$2\to 2$ with various $M_{Zq}$ thresholds
(discontinuous lines, as indicated).
Each curve with a defininte $M_{Zq}$ threshold 
value stops when the curve of the next threshold value takes over. 
Also shown with open circles is 
the signal due to an excited quark of 500 GeV mass and narrow width (\scriptR\  $= 1$).
The resonance production cross section is taken from Ref. \cite{Baur90}.}
\end{figure}

In Fig.~\ref{fig:MZj1DaMCcompnocut} we compare the $M_{Zj1}$
distribution of the data with the {\sc pythia} $2\to 1$
process and with the {\sc pythia} $2\to 2$ processes
with various $M_{Zq}$ thresholds. For the $2\to 1$
process, the MC is normalized to the total number of 
data events. A different but common normalization factor is
used for all $2\to 2$ processes determined using the 
$M_{Zq}>100$ GeV MC sample for $M_{Zj1}>150$ GeV.
The $2\to 1$
simulation agrees well with the data but provides
sufficient statistics only for $M_{Zj1}<300$ GeV.
On the other hand, the $2\to 2$ processes describe
the data with reasonable precision for $M_{Zj1}>150$ GeV. 
Since the latter is the region of interest for the present search, 
we have used only the $2\to 2$ process for 
estimation of the SM background with an $M_{Zq}$ threshold
chosen according to the $M_{Zj1}$ region to be investigated.
Also shown in Fig.~\ref{fig:MZj1DaMCcompnocut} is
the signal due to an excited quark of 500 GeV mass and narrow width (\scriptR\  $= 1$).

\begin{figure}[h]
\includegraphics[scale=0.4]{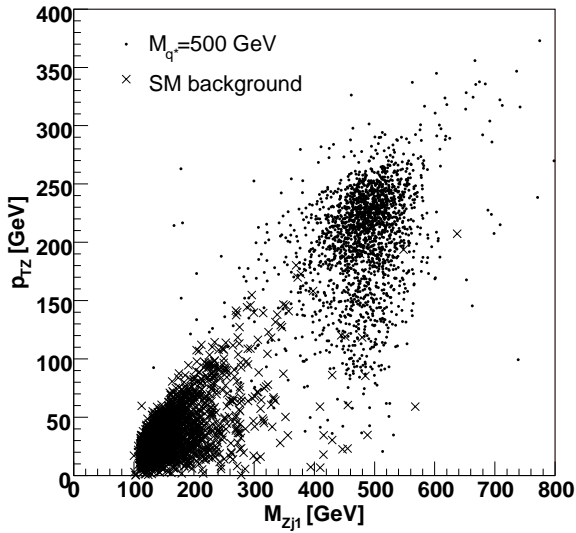}
\caption{ \label{fig:pTZvsMZj} $p_{TZ}$ vs $M_{Zj1}$ distributions for a resonance of
mass of 500 GeV ($\xi = 1$) and for the SM background.  
Both the signal and background events passed through complete reconstruction.
Each distribution is arbitrarily normalized.
}
\end{figure}
 
Since no significant excess of events is observed,
which would indicate the presence of a resonance, 
we determined the upper limit
on the production cross section of a hypothetical resonance as a function of
its mass and width.
We made use of the fact that in the $p_{TZ}$ vs $M_{Zj1}$ plane,
events from the resonance are concentrated for $M_{Zj1}$ around the mass value and 
for $p_{TZ}$ at about half of the mass value
of the resonance, since the resonance is nearly at rest.
The SM background does not exhibit a similar structure,
as it is shown in Fig.~\ref{fig:pTZvsMZj}. 
In addition, finite width and mass resolutions wash out
the correlation between  $p_{TZ}$ and $M_{Zj1}$.
We therefore considered events around  the peak values $M_{Zj1}^c$ and  
$p_{TZ}^c$ of the resonance
determined by the following condition:
\begin{equation}\label{eq:topocut}
\left( \frac{M_{Zj1}-M_{Zj1}^c}{M_{Zj1}^{rms}} \right)^2 + 
\left( \frac{p_{TZ}-p_{TZ}^c}{p_{TZ}^{rms}} \right)^2 < k^2
\end{equation}
and we optimized the cut value $k$.
Here, $M_{Zj1}^{rms}$ and $p_{TZ}^{rms}$ are the RMS values 
of the corresponding distributions of the resonance.
At given values of mass and width, the latter defined by \scriptR , we 
varied $k$ in Eqn.(\ref{eq:topocut}) 
between $0$ and $3$ in steps of $0.1$. 
\begin{table}[h]
\caption{\label{tab:results} Measured ($\sigma_{95}$) and expected ($\sigma_{95}^{ave}$)
values of the upper limit on the resonance cross section  times  branching fraction,
signal acceptance $\times$ efficiency, SM background, and number of observed 
events at the optimal value of the 
topological cut $k$ for different resonance masses and for \scriptR\ $=1$.}
\begin{ruledtabular}
\begin{tabular}{ccccccc}
 $M_{q^*}$  & $k$  & $\sigma_{95}$  & $\sigma_{95}^{ave}$  & Acceptance & SM              & Data  \\
   (GeV)      &      &   (pb)         &     (pb)        &    $\times$ efficiency        & background      &   (events)    \\
\hline
  300 &   1.1 & 0.25 &  0.290 &  0.140 $\pm$ 0.009 &   32.8 $\pm$   2.9 &  31    \\ 
  400 &   1.2 & 0.15 &  0.129 &  0.164 $\pm$ 0.010 &   7.5  $\pm$   0.8 &   9    \\ 
  500 &   1.3 & 0.08 &  0.079 &  0.195 $\pm$ 0.012 &   2.9  $\pm$   0.8 &   3    \\
  600 &   1.8 & 0.05 &  0.053 &  0.244 $\pm$ 0.014 &   1.6  $\pm$   0.6 &   1    \\
  700 &   1.7 & 0.03 &  0.044 &  0.243 $\pm$ 0.014 &   0.64 $\pm$   0.06 &   0    \\
\end{tabular}
\end{ruledtabular}
\end{table}
\begin{table}[h]
\caption{\label{tab:sigmas} Measured upper limit on the resonance
cross section times  branching fraction 
at the 95\% C.L., $\sigma_{95}$, for different resonance masses and \scriptR\ values.
$\sigma_{q^*}$, the production cross section of an excited quark
times its decay branching fraction into $Z+$jet and $Z\to e^+e^-$, is
calculated in LO. The width for \scriptR\ $=1$ is also shown \cite{Baur90} 
for each resonance mass.
Cross sections are quoted in pb, whereas masses and widths are in GeV.}
\begin{ruledtabular}
\begin{tabular}{ccccc|cc}
 Mass  &  \multicolumn{4}{c|}{$\sigma_{95}$ } & $\sigma_{q^*}$ & $\Gamma_{q^*}$ \\

 &  \scriptR\  $=0.3$ & 0.5 & 0.7 &  1 & \multicolumn{2}{c}{\scriptR\  $=1$}   \\ \hline
300 &       &       &       & 0.25 &  2.045 & 13 \\
400 & 0.32 & 0.16 & 0.15 & 0.15 &  0.382 & 16 \\
500 & 0.17 & 0.08 & 0.07 & 0.08 &  0.084 & 20 \\
600 & 0.10 & 0.06 & 0.05 & 0.05 &  0.021 & 24 \\
700 & 0.07 & 0.05 & 0.05 & 0.03 &  0.005 & 27 \\
\end{tabular}
\end{ruledtabular}
\end{table}
\begin{figure}[h]
\includegraphics[scale=0.4]{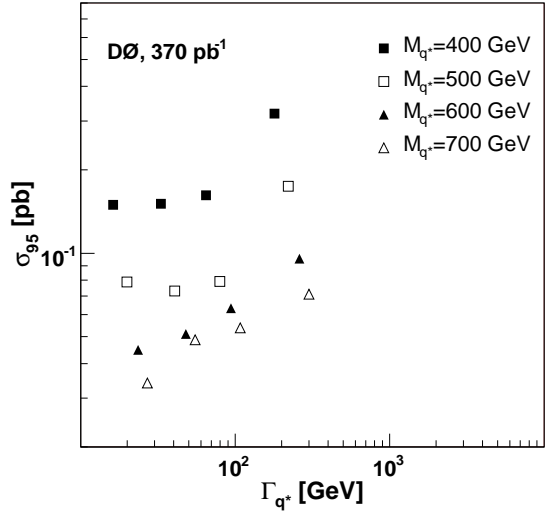}
\caption{ \label{fig:Xsection95vsWidth} Upper limit on the resonance
cross section 
times  branching fraction
at the 95\% C.L., $\sigma_{95}$, for different resonance masses as a function of the resonance width.
}
\end{figure}
\begin{figure}[h]
\includegraphics[scale=0.4]{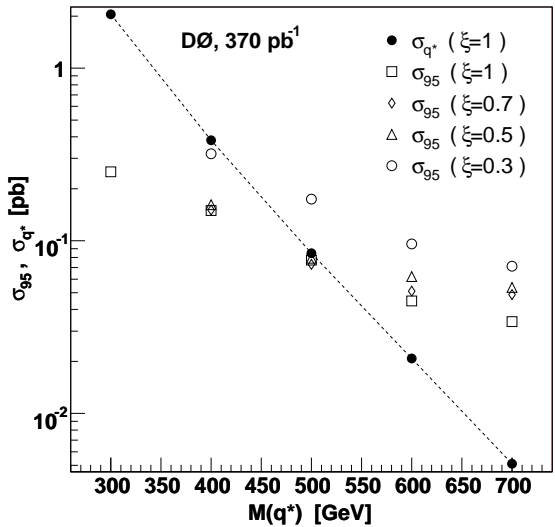}
\caption{ \label{fig:Xsection95ElliptCut} 
Upper limits on the resonance cross section 
times  branching fraction
at the 95\% C.L., $\sigma_{95}$, for different \scriptR\ values as functions of the resonance mass
(open symbols).
Full circles indicate the LO production cross section
of an excited quark times its decay branching fraction into $Z+$jet and
$Z\to e^+e^-$, $\sigma_{q*}$, for \scriptR\ $=1$ \cite{Baur90}. 
}
\end{figure}

Based only on the information from the signal and background simulation, for each $k$ 
we calculated 
$\sigma_{95}^{ave}$, the {\it expected value} of the upper limit on the resonance production
cross section times  branching fraction at the 95\% C.L. using a Bayesian approach \cite{d0limit}
and by averaging over possible outcomes of the background-only hypothesis assuming
Poisson statistics of the background.
The optimum value of $k$ corresponds to the minimum value of $\sigma_{95}^{ave}$.
At this value of $k$, using also the data, we derived $\sigma_{95}$, 
the {\it measured value} of the upper limit on the resonance production
cross section times  branching fraction at the 95\% C.L.
In this calculation we have taken into account systematic uncertainties
in the determination of the luminosity (6.5\%), trigger and identification efficiencies,
and those of the jet calibration and resolution. Systematic uncertainties
due to the modeling of the SM background and to the choice of the PDF, 
as well as those due to the threshold of the $M_{Zj1}>150$ GeV in the normalization of the
background, have also been included. 

In Table~\ref{tab:results},
$\sigma_{95}$ and $\sigma_{95}^{ave}$
are shown together with the signal acceptance, the
SM background level, and the number of data events for \scriptR\  $=1$.
The measured  $\sigma_{95}$ values are displayed in Fig.~\ref{fig:Xsection95vsWidth}
and Fig.~\ref{fig:Xsection95ElliptCut}, 
and are compiled in Table~\ref{tab:sigmas} for different masses and widths (\scriptR ). 
In Fig.~\ref{fig:Xsection95ElliptCut}, 
also shown is $\sigma_{q*}$, the LO production cross section
of an excited quark times its decay branching fraction into $Z+$jet and
$Z\to e^+e^-$, for \scriptR\ $=1$  \cite{Baur90}. 
We find a lower limit of 510 GeV at the 95\% C.L. 
for the mass of an excited quark 
for \scriptR\  $=1$ 
within the framework of the model considered.
In earlier measurements, lower bounds of 460, 530 and 775 GeV
were obtained for the same quantity,
but in different decay modes, namely 
in $q^*\to q\gamma$ \cite{CDFWgamj}, $q^*\to qW$ \cite{CDFWgamj}, and
$q^*\to qg$ \cite{D0dij}, respectively, and therefore with
different systematics. 

In conclusion, we have searched for a resonance produced by the fusion of a gluon and
a quark in $p\bar p$ collisions at a center of mass energy of 1.96 TeV which
decays into a $Z$ boson and a quark in the $Z\to e^+e^-$ decay channel. 
In the absence of a signal, we have determined 95\% C.L. upper limits on the cross section 
times  branching fraction as a function of the mass and width of the resonance.
The present study is complementary to earlier searches because
it has sensitivity to hypothetical models
with enhanced couplings to the $Z$ boson.


%
We thank the staffs at Fermilab and collaborating institutions, 
and acknowledge support from the 
DOE and NSF (USA);
CEA and CNRS/IN2P3 (France);
FASI, Rosatom and RFBR (Russia);
CAPES, CNPq, FAPERJ, FAPESP and FUNDUNESP (Brazil);
DAE and DST (India);
Colciencias (Colombia);
CONACyT (Mexico);
KRF and KOSEF (Korea);
CONICET and UBACyT (Argentina);
FOM (The Netherlands);
PPARC (United Kingdom);
MSMT (Czech Republic);
CRC Program, CFI, NSERC and WestGrid Project (Canada);
BMBF and DFG (Germany);
SFI (Ireland);
The Swedish Research Council (Sweden);
Research Corporation;
Alexander von Humboldt Foundation;
and the Marie Curie Program.
%

\end{document}